\documentclass{elsart}

\usepackage{harvard}

\input{psfig}

\usepackage{amssymb}

\def\url#1{{\ttfamily\def\/{/\discretionary{}{}{}}#1}}

\begin{document}

\begin{frontmatter}
\title{Introduction to Microwave Background Polarization}
\author{Arthur Kosowsky\thanksref{email}}
\address{
Department of Physics and Astronomy, Rutgers University, 136 Frelinghuysen Road, Piscataway, NJ 08854-8019 USA
}
\thanks[email]{Email: kosowsky@physics.rutgers.edu}


\end{frontmatter}





\def\hatn{{\bf \hat n}}

\section{Introduction}
\label{intro}

Through the hard work of many excellent experimenters, the
detection and characterization of temperature fluctuations
in the cosmic microwave background is now on sure footing.
Following the initial detection by the DMR instrument
aboard COBE \cite{Smoetal90}, numerous ground and balloon-based
detections have been made, and the first reasonably large
temperature maps at angular scales of a degree
have been constructed \cite{Devetal98}. The upcoming satellites
MAP and Planck promise full sky temperature maps of
unprecedented resolution and sensitivity, as detailed
elsewhere in these lectures. Theoretically, much of this
intensive effort has been motivated by the realization
that the microwave background
temperature fluctuations contain a wealth of 
fundamental cosmological information 
\cite{Junetal96,ZalSelSpe97,EisHuTeg99}.

Polarization of the microwave background is a different
story. Polarization is expected in every cosmological
model, for the simple reason that the Thomson scattering
which thermalizes the radiation has a polarization-dependent
cross section. But the polarization signal is generically
expected to be a factor of 10 to 50 smaller than the
temperature fluctuations, presenting that much greater
of an experimental challenge. Only upper limits on
polarization of around a part in $10^5$ now exist, but a
new generation of experiments optimized for polarization
are currently being constructed, which potentially
have both the raw sensitivity
and the control over systematic errors
necessary to make the first detection.
In many ways, the 
experimental study of polarization today is
at about the same stage that temperature was ten years
ago.

This contribution aims to explain how polarization is
physically characterized, 
how it is generated in the microwave background,
the mathematical description of the associated
power spectra, and the physical effects which might
be probed via polarization measurements. Another
elementary reference about microwave background
polarization from a somewhat different
perspective is \citeasnoun{WhiHu97}.

\section{Review of Stokes Parameters}
\label{stokes}

Polarized light is conventionally described in terms of the
Stokes parameters, which are presented in any optics text.
Consider a
nearly monochromatic plane electromagnetic wave propogating in
the $z$-direction; nearly monochromatic here means that its frequency
components are closely distributed around its mean frequency
$\omega_0$. 
The components of the wave's electric field vector at a given
point in space can be written as 
\begin{equation}
E_x=a_x(t)\cos\left[\omega_0 t - \theta_x(t)\right],\quad
E_y=a_y(t)\cos\left[\omega_0 t - \theta_y(t)\right].
\label{efield}
\end{equation}
The requirement that the wave is nearly monochromatic guarantees
that the amplitudes $a_x$ and $a_y$ and the phase angles
$\theta_x$ and $\theta_y$ will vary slowly
relative to the inverse frequency of the wave. If some
correlation exists between the two components in
Eq.~(\ref{efield}), then the wave is polarized.

The Stokes parameters are defined as the following time
averages: 
\label{stokesdef}
\begin{eqnarray}
I\,&&\equiv \langle a_x^2\rangle + \langle a_y^2\rangle;\\
Q\,&&\equiv \langle a_x^2\rangle - \langle a_y^2\rangle;\\
U\,&&\equiv \langle 2a_xa_y\cos(\theta_x -\theta_y)\rangle;\\
V\,&&\equiv \langle 2a_xa_y\sin(\theta_x -\theta_y)\rangle.
\end{eqnarray}
The averages are over times long compared to the inverse
frequency of the wave.
The parameter $I$ gives the intensity of the radiation which is
always positive and is equivalent to the temperature for
blackbody radiation. The other three parameters define the
polarization state of the wave and can have either sign. 
Unpolarized radiation, or ``natural light,'' is described by
$Q=U=V=0$. 

The parameters $I$ and $V$ are physical observables independent
of the coordinate system, but $Q$ and $U$ depend on the
orientation of the $x$ and $y$ axes. If a given wave is
described by the parameters $Q$ and $U$ for a certain
orientation of the coordinate system, then after a rotation of
the $x-y$ plane through an angle $\phi$, it is
straightforward to verify that the same wave is now
described by the parameters
\begin{eqnarray}
Q'&&=Q\cos(2\phi) + U\sin(2\phi),\nonumber\\
U'&&=-Q\sin(2\phi) + U\cos(2\phi).
\label{uqtransf}
\end{eqnarray}
From this transformation it is easy to see that the quantity
$Q^2+U^2$ is invariant under rotation of the axes, and the angle
\begin{equation}
\alpha\equiv{1\over 2}\tan^{-1}{U\over Q}
\label{alpha}
\end{equation}
transforms to $\alpha -\phi$ under a rotation by $\phi$ and thus
defines a constant orientation, which physically is parallel
to the electric field of the wave.
The Stokes parameters are a useful description of
polarization because they are {\it additive} for incoherent
superposition of radiation; note this is not true for the
magnitude or orientation of polarization.

While polarization has a magnitude and an orientation, it
is not a vector quantity because the orientation does not
have a direction, describing only the plane in which the
electric field of the wave oscillates. Mathematically,
the Stokes parameters are identical for an axis rotation
through an angle of $\pi$, whereas for a vector, such a
rotation would lead to an inverted vector and a full rotation
through $2\pi$ is required to return to the same situation.
The transformation law in Eq.~(\ref{uqtransf}) is characteristic
of the second-rank {\it tensor}
\begin{equation}
\rho={1\over 2} \pmatrix{\vphantom\int I+Q & U-iV\cr 
\vphantom\int U+iV& I-Q},
\label{densmatrix}
\end{equation}
which also corresponds to the quantum mechanical
density matrix for an ensemble of photons \cite{Kos96}
(the matrix is 2 by 2
because the photon has two helicity states).

\section{Polarization from Thomson Scattering}
\label{polgeneration}

Polarization in the microwave background is generated through
the polarization-dependent cross-section for Thomson scattering.
Consider Thomson scattering of an incoming unpolarized beam of
electromagnetic radiation by an electron; this discussion closely
follows those in \citeasnoun{Kos96} and \citeasnoun{Kos98}. 
The total scattering
cross-section, defined as the radiated intensity per unit solid
angle divided by the incoming intensity per unit area, is given by
\begin{equation}
{d\sigma\over d\Omega} = {3\sigma_T\over 8\pi} 
\left|{\hat\varepsilon}'\cdot{\hat\varepsilon}\right|^2
\label{thomson}
\end{equation}
where $\sigma_T$ is the total Thomson cross section and
the vectors $\hat\varepsilon$ and ${\hat\varepsilon}'$ are unit
vectors in the planes perpendicular to the propogation directions
which are aligned with the outgoing and incoming polarization,
respectively. Consider first a nearly monochromatic, unpolarized 
incident plane wave of intensity $I'$ and cross-sectional area $\sigma_B$
which is scattered into the $z$-axis direction.
Defining the $y$-axes of the incoming and outgoing coordinate
systems to be in the scattering plane, the Stokes parameters
of the outgoing beam, defined with respect to the $x$-axis,
follow from Eq.~(\ref{thomson}) as 
\begin{eqnarray}
I &=& {3\sigma_T\over 8\pi\sigma_B}I'(1+\cos^2\theta),\\
Q &=& {3\sigma_T\over 8\pi\sigma_B}I'\sin^2\theta,\\
U &=& 0,
\end{eqnarray}
\label{stokesout}
where $\theta$ is the angle between the incoming and outgoing beams.
By symmetry, Thomson scattering can generate no circular polarization,
so $V=0$ always and will not be considered further.
(Note that Eqs.~(\ref{stokesout}) give the well-known result that 
sunlight from the horizon at midday is linearly polarized parallel
to the horizon).

The net polarization produced by the scattering of an incoming,
unpolarized ratiation field of intensity $I'(\theta,\phi)$ is determined
by integrating Eqs.~(\ref{stokesout}) over all incoming directions. Note
that the coordinate system for each incoming direction
must be rotated about the $z$-axis so that the outgoing Stokes parameters
are all defined with respect to  a common coordinate system, using the
transformation of $Q$ and $U$ under rotations. The result is
\begin{eqnarray}
I({\bf\hat z}) &=& {3\sigma_T\over 16\pi\sigma_B}\int d\Omega 
(1+\cos^2\theta) I'(\theta,\phi),\\
Q({\bf\hat z}) - iU({\bf\hat z}) 
&=& {3\sigma_T\over 16\pi\sigma_B}\int d\Omega 
\sin^2\theta e^{2i\phi} I'(\theta,\phi).
\end{eqnarray}
\label{stokesint}
Expanding the incident radiation field in spherical harmonics,
\begin{equation}
I'(\theta,\phi) = \sum_{lm}a_{lm}Y_{lm}(\theta,\phi),
\end{equation}
leads to the following expressions for the outgoing Stokes parameters:
\begin{eqnarray}
I({\bf\hat z})&=&{3\sigma_T\over 16\pi\sigma_B}
\left[{8\over 3}\sqrt{\pi}\, a_{00} + {4\over 3}\sqrt{\pi\over 5} a_{20}
\right],\\
Q({\bf\hat z}) - iU({\bf\hat z}) &=&
{3\sigma_T\over 4\pi\sigma_B} \sqrt{2\pi\over 15} a_{22}.
\end{eqnarray}
\label{a22}
Thus polarization is generated along the outgoing $z$-axis provided that
the $a_{22}$ quadrupole moment of the incoming radiation is non-zero. To
determine the outgoing polarization in a direction making an angle $\beta$
with the $z$-axis, the same physical incoming field 
must be multipole expanded
in a coordinate system rotated through the Euler angle $\beta$; 
the rotated
multipole coefficients are
\begin{eqnarray}
{\tilde a}_{lm} &=& \int d\Omega Y^*_{lm}(R\Omega) I'(\Omega)\nonumber\\
&=&\sum_{m'=-m}^m D^{l\,*}_{m'm}(R)\int d\Omega Y^*_{lm'}(\Omega)
I'(\Omega),
\label{almtilde}
\end{eqnarray}
where $R$ is the rotation operator and $D^l_{m'm}$ is the Wigner
D-symbol. (For a wonderfully complete reference on representations
of the rotation group, see \citeasnoun{varshalovich}).
In the rotated coordinate system,
the multipole coefficient generating polarization is ${\tilde a}_{22}$ by
Eqs.~(\ref{a22}). The unrotated multipole components which contribute
to polarization will all have $l=2$ by the orthogonality of the
spherical harmonics. If the incoming radiation field is independent of
$\phi$, as it will be for individual Fourier components of a density
perturbation, then
\begin{equation}
{\tilde a}_{22} = a_{20} d^{2\, *}_{02}(\beta) 
={\sqrt{6}\over 4}a_{20}\sin^2\beta,
\label{a20}
\end{equation}
which has used an explicit expression for the reduced D-symbol
$d^l_{m'm}$. The outgoing Stokes parameters are finally
\begin{equation}
Q({\bf\hat n})-iU({\bf\hat n}) 
={3\sigma_T\over 8\pi\sigma_B} \sqrt{\pi\over 5}\,a_{20}\sin^2\beta.
\label{a20}
\end{equation}
In other words, an azimuthally-symmetric
radiation field will generate a polarized scattered field if
it has a non-zero $a_{20}$ multipole component, and the magnitude
of the scattered polarization will be proportional to $\sin^2\beta$.
Since the incoming field is real, $a_{20}$ will be real, $U=0$,
and the polarization orientation will be in the plane of the $z$-axis
and the scattering direction. Similar relationships can be derived
for radiation fields which are not azimuthally symmetric, which
occur in the cases of vector and tensor metric perturbations.

In short, what this section shows is that unpolarized quadrupolar
radiation fields get Thomson scattered into polarized radiation
fields. This is the key fact which must be appreciated to understand
why the microwave background should be polarized and what the
magnitude of the polarization is expected to be.

\section{Generation of Microwave Background Polarization}
\label{generation}

At times significantly before decoupling, the universe is
hot enough that protons and electrons exist freely in a plasma.
During this epoch, the rate for photons to Thomson scatter
off of free electrons is large compared to the expansion
rate of the universe. Thus, the photons and electrons stay
in thermal equilibrium at a common temperature and are
said to be tightly coupled. As the universe drops below
a temperature of around 0.1 eV at a redshift
of around 1300, the electrons and protons begin to ``recombine''
into neutral hydrogen. Within a short time, almost all
the free electrons are converted to neutral hydrogen,
the rapid Thomson scattering ceases for lack of scatterers,
and the radiation is said to decouple. At this point, the radiation
will propogate freely until the universe reionizes at some
redshift greater than 5.

During the tight coupling epoch, the photons must have a distribution
which mirrors that of the electrons. An immediate consequence is
that the angular dependence of the radiation field at a given point
can only possess a monopole (corresponding to the temperature)
and a dipole (corresponding to a Doppler shift from a peculiar
velocity) component, and that the radiation field is
unpolarized. Any higher multipole moment will rapidly
damp away as the electrons scatter off the free electrons,
and no net polarization can be produced through scattering.

A quadrupole is subsequently produced at decoupling
as free streaming of the photons begins. A single Fourier mode of the
radiation field can be described by the temperature distribution
function $\Theta(k,\mu,\eta)$ where $k$ is the wavenumber, 
$\mu = {\bf\hat k}\cdot{\bf\hat n}$ is the angle between the 
vector $\bf k$ and the propagation direction $\bf\hat n$, 
and $\eta$ is conformal time. (For mathematical simplicity only a flat
universe is considered here, although the non-flat cases are
no more complicated conceptually.) Ignoring
gravitational potential contributions, free streaming of the photons
is described by the Liouville equation
$\dot\Theta + ik\mu\Theta = 0$. If the free streaming begins at
time $\eta_*$, then the solution at a later time is simply
$\Theta(k,\mu,\eta) = \Theta(k,\mu,\eta_*)\exp(-ik\mu(\eta - \eta_*))$.
We can reexpress the $\mu$ dependence as a multipole expansion
\begin{equation}
\Theta(k,\mu,\eta) = \sum_{l=0}^\infty (-i)^l\Theta_l(k,\eta)P_l(\mu);
\end{equation}
using the identity
\begin{equation}
e^{iz\cos\phi}=\sum_{n=0}^\infty (2n+1)i^n j_n(z)P_n(\cos\phi)
\end{equation}
the free streaming becomes
\begin{equation}
\Theta_l(k,\eta) = (2l+1)[
\Theta_0(k,\eta_*)j_l(k\eta - k\eta_*)
+ \Theta_1(k,\eta_*)j'_l(k\eta - k\eta_*)],
\label{freestream}
\end{equation}
where $j_l$ is the usual spherical
Bessel function.

We are interested in the behavior of the free streaming 
at times near decoupling; at later times, the number density of
free electrons which can Thomson scatter has dropped to negligible
levels and no further polarization can be produced. The physical
length scales of interest for microwave background fluctuations
will be larger than the thickness of the last scattering surface,
so $k(\eta-\eta_*)$ will be small compared to unity. For small
arguments $x\ll 1$, $j_l(x)/j_l'(x)\sim x/l$, 
which implies that if the monopole
and dipole radiation components are initially of comparable size,
free streaming through the region of polarization generation 
with thickness $\Delta$ will generate
a quadrupole component from the dipole which is a factor of $2/(k\Delta)$
larger than the quadrupole component from the monopole. In other
words, on length scales large compared to the thickness of 
the surface of last scattering, the quadrupole moment and thus
the polarization couples much more strongly to the velocity
of the baryon-photon fluid than to the density.
Note that on smaller scales with $k\Delta\gtrsim 1$, the polarization
can couple more strongly to either the velocity or the density,
depending on the scale, but for standard recombination
these scales are always small enough
that the microwave background fluctuations are strongly
diffusion damped.

\section{Polarization and Sound Waves}
\label{acoustic}

Inflation produces acoustic oscillations in the early
universe which are {\it coherent}: all Fourier modes of
a given wavelength have the same phase. Such acoustic
oscillations have a very specific relationship
between velocity and density perturbations, which
shows up in the relative angular scales of features
in the temperature and polarization power spectra.
As emphasized in \citeasnoun{HuSug96},
the photon-baryon density perturbation in the tight-coupling
regime obeys the differential equation for a forced, damped
harmonic oscillator with the damping coming from the expansion
of the universe and the forcing from gravitational potential
perturbations. The solution is of the form
\begin{equation}
\Theta_0(k,\eta) =  A_1(\eta)\cos(kr_s) + A_2(\eta)\sin(kr_s)
\label{theta0}
\end{equation}
where the amplitudes vary slowly in time and $r_s\simeq \eta/\sqrt{3}$
is the sound horizon.
The velocity perturbation follows from the photon continuity
equation $\dot\Theta_0 = -k\Theta_1/3$, again neglecting gravitational
potential perturbations. A detailed consideration of boundary
conditions reveals that initial isentropic density perturbations couple
to the cosine harmonic in the small-scale limit, and this approximation
is good even for the largest-wavelength acoustic oscillations 
\cite{HuWhi96}.
Thus in an inflationary model, at the surface of last scattering,
the photon monopole has a $k$-dependence of approximately
$\cos(k\eta_*/\sqrt{3})$,
while the dipole, which is the main contributor to the polarization,
has a $k$-dependence of approximately $\sin(k\eta_*/\sqrt{3})$.
For initial isocurvature perturbations, the density perturbations
couple instead to the sine harmonic, but the photon monopole
and dipole are still $\pi/2$ out of phase.

\begin{figure}[htbp] 
\centerline{\psfig{file=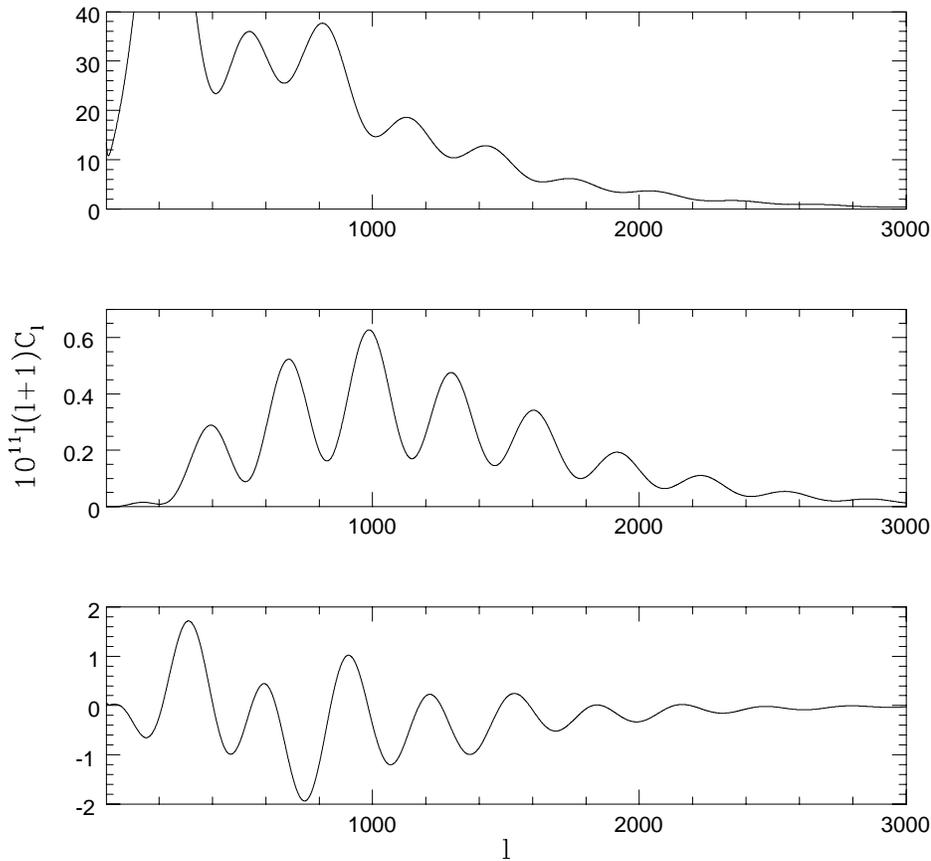,width=5in}}
\caption{The power spectra for temperature fluctuations (top),
polarization (center) and temperature-polarization cross-correlation
(bottom) for a typical inflationary model. The oscillations remain
in phase up to $l=3000$.
}
\label{fig:interleave}
\end{figure}

Squaring these amplitudes gives the rough behavior of the
CMB power spectra. Acoustic peaks in the temperature power
spectrum occur at scales where $\cos^2(k\eta_*/\sqrt{3})$ 
has its maxima. The amplitude of the velocity perturbations 
are suppressed
by a factor of $c_s$ with respect to the density perturbations,
so the temperature peaks reflect only the density perturbations.
The
polarization couples to the temperature dipole on scales larger
than the thickness of the last scattering surface, and acoustic
peaks in the polarization power spectrum will be present at scales
where $\sin^2(k\eta_*/\sqrt{3})$ has its maxima. In other words,
the temperature peaks represent density extrema, the polarization
peaks represent velocity extrema, and for coherent oscillations
these two sets of maxima are at {\it interleaved angular
scales} (see Fig.~\ref{fig:interleave}). 
This is a generic signature of coherent acoustic
oscillations and is likely the most easily measurable
physics signal in microwave background polarization.
If two peaks are detected in the temperature power spectrum,
the angular scale between the two makes a tempting
target for polarization measurements.

The cross-correlation
between the temperature and polarization will have extrema as
$-\cos(k\eta_*/\sqrt{3})\sin(k\eta_*/\sqrt{3})$ which fall
between the temperature and polarization peaks. (The correlation
between the polarization and the velocity contribution to the temperature
averages to zero because of their different angular dependences.)
The sign of the cross-correlation peaks can be used to
deduce whether a temperature peak represents a compression or a
rarefaction, which can be checked against the alternating
peak-height signature
if the universe has a large enough baryon fraction \cite{HuSug96}.

A combination of isentropic and isocurvature fluctuations shifts
all acoustic phases by the same amount if the ratio of their 
amplitudes is independent of scale, thus leaving the acoustic
signature intact. If the amplitude ratio depends on scale, the coherent
acoustic oscillations could be modified, but fine tuning would be 
required to wash them out completely. Multi-field inflation models
generically produce both isocurvature and isentropic perturbations
\cite{KofLin87,MukSte98} 
but the resulting microwave background power spectra
are just beginning to be studied in detail \cite{Kanetal98}.

\section{The Tensor Harmonic Expansion}
\label{tensor}

The last two sections have pulled a fast one. We began
by discussing polarization as a two component tensor quantity,
but then started discussing the production of polarization as
if only its amplitude were relevant. A more complete
formalism for describing the polarization field
has been worked out and will be presented in this section
(see \citeasnoun{KamKosSte97} for a more extensive discussion).
An equivalent formalism employing spin-weighted spherical
harmonics has been used extensively by \citeasnoun{ZalSel97}. 
Note that the normalizations employed by Seljak
and Zaldarriaga are slightly different than those
adopted here and by \citeasnoun{KamKosSte97}.


The microwave background temperature pattern on the sky
$T(\hatn)$ is conventionally expanded in a
complete set of orthonormal basis functions, the spherical harmonics:
\begin{equation}
     {T(\hatn) \over T_0}=1+\sum_{l=1}^\infty\sum_{m=-l}^l
     a^{\rm T}_{(lm)}\,Y_{(lm)}(\hatn)
\label{Texpansion}
\end{equation}
where
\begin{equation}
 a^{\rm T}_{(lm)}={1\over T_0}\int d\hatn\,T(\hatn) Y_{(lm)}^*(\hatn)
\label{temperaturemoments}
\end{equation}
are the temperature multipole coefficients and $T_0$ is the
mean CMB temperature. 
Similarly, we can expand the polarization tensor for linear
polarization, 
\begin{equation}
  P_{ab}(\hatn)={1\over 2} \left( \begin{array}{cc}
   Q(\hatn) & -U(\hatn) \sin\theta \\
   \noalign{\vskip6pt}
   - U(\hatn)\sin\theta & -Q(\hatn)\sin^2\theta \\
   \end{array} \right)
\label{whatPis}
\end{equation}
(compare with Eq.~\ref{densmatrix}; the extra factors are
convenient
because the usual spherical coordinate basis is orthogonal
but not orthonormal) in terms of 
{\it tensor spherical harmonics}, a complete set of
orthonormal basis functions for symmetric trace-free $2\times2$ tensors 
on the sky, 
\begin{equation}
     {P_{ab}(\hatn)\over T_0} = \sum_{l=2}^\infty\sum_{m=-l}^l
     \left[ a_{(lm)}^{\rm G}
     Y_{(lm)ab}^{\rm G}(\hatn) + a_{(lm)}^{\rm C} Y_{(lm)ab}^{\rm C}
     (\hatn) \right],
\label{Pexpansion}
\end{equation}
where the expansion coefficients are given by
\begin{eqnarray}
     a_{(lm)}^{\rm G}&=&{1\over T_0}\int \, d\hatn P_{ab}(\hatn)
               Y_{(lm)}^{{\rm G} \,ab\, *}(\hatn), \\
     a_{(lm)}^{\rm C}&=&{1\over T_0}\int d\hatn\, P_{ab}(\hatn)
                             Y_{(lm)}^{{\rm C} \, ab\, *}(\hatn),
\label{defmoments}
\end{eqnarray}
which follow from the orthonormality properties
\begin{equation}
 \int d\hatn\,Y_{(lm)ab}^{{\rm G}\,*}(\hatn)\,
       Y_{(l'm')}^{{\rm G}\,\,ab}(\hatn)
=\int d\hatn\,Y_{(lm)ab}^{{\rm C}\,*}(\hatn)\,
       Y_{(l'm')}^{{\rm C}\,\,ab}(\hatn)
=\delta_{ll'} \delta_{mm'},\nonumber
\end{equation}
\begin{equation}
\int d\hatn\,Y_{(lm)ab}^{{\rm G}\, *}(\hatn)\,
Y_{(l'm')}^{{\rm C}\,\,ab}(\hatn)
=0.
\label{norms}
\end{equation}

These tensor spherical harmonics have been used
primarily in the literature of gravitational
radiation, where the metric perturbation can be
expanded in these tensors. Explicit forms can be
derived via various algebraic and group theoretic
methods; see \citeasnoun{Thorne} for a complete discussion.
A particularly elegant and useful derivation
of the tensor spherical harmonics (along with
the vector spherical harmonics as well) is provided
by differential geometry \cite{Ste96}. Given a scalar
function on a manifold, the only related vector quantity
at a given point of the manifold is the covariant
derivative of the scalar function. 
The tensor basis functions can be
derived by taking the scalar basis functions $Y_{lm}$
and applying to them two covariant derivative operators
on the manifold of the two-sphere (the sky):
\begin{equation}
     Y_{(lm)ab}^{\rm G} = N_l
     \left( Y_{(lm):ab} - {1\over2} g_{ab} Y_{(lm):c}{}^c \right),
\label{Yplusdefn}
\end{equation}
and
\begin{equation}
     Y_{(lm)ab}^{\rm C} = { N_l \over 2}    
     \left(\vphantom{1\over 2} 
       Y_{(lm):ac} \epsilon^c{}_b +Y_{(lm):bc} \epsilon^c{}_a \right),
\label{Ytimesdefn}
\end{equation}
where $\epsilon_{ab}$ is the completely antisymmetric tensor,
the ``:'' denotes covariant differentiation on the 2-sphere,
and
\begin{equation}
     N_l \equiv \sqrt{ {2 (l-2)! \over (l+2)!}}
\label{Nleqn}
\end{equation}
is a normalization factor. Note that the somewhat more
familiar vector spherical harmonics used to describe
electromagnetic multipole radiation 
can likewise be derived as a
single covariant derivative of the scalar spherical harmonics.

While the formalism of differential geometry may look imposing
at first glance, the expansion of the polarization
field has been cast into exactly the same form as for the familiar
temperature case, with only the extra complication of evaluating
covariant derivatives.
Explicit forms for the tensor harmonics are given in 
\citeasnoun{KamKosSte97}.
Note that the underlying manifold, the two-sphere, is the
simplest non-trivial manifold, with a constant Ricci curvature
$R=2$, so the differential geometry is easy.
One particularly useful property for doing calculations
is that the covariant derivatives are subject to integration
by parts:
\begin{equation}
\int d{\bf\hat n}AB_{:a} = -\int d{\bf\hat n}A_{:a}B
\end{equation}
with no surface term if the integral is over the entire sky.
Also, the scalar spherical harmonics are eigenvalues
of the Laplacian operator:
\begin{equation}
Y_{(lm):a}{}^{:a}\equiv \nabla^2 Y_{(lm)} = -l(l+1)Y_{(lm)}.
\end{equation}

The existence of two sets of basis functions,
labeled here by ``G'' and
``C'', is due to the fact that the symmetric traceless
$2\times2$ tensor describing linear polarization
is specified by two independent parameters.
In two dimensions, any symmetric traceless
tensor can be uniquely decomposed into a
part of the form $A_{:ab}-(1/2)g_{ab} A_{:c}{}^c$ 
and another part of the form
$B_{:ac}\epsilon^c{}_b+B_{:bc}\epsilon^c{}_a$ 
where $A$ and $B$ are two scalar
functions.   This decomposition is quite similar to the decomposition of a
vector field into a part which is the gradient of a scalar field 
and a part which is the curl of a vector field; hence we use the notation
G for ``gradient'' and C for ``curl''. In fact, this correspondence is
more than just cosmetic: if a linear polarization field is visualized
in the usual way with headless ``vectors'' representing the amplitude
and orientation of the polarization, then
the G harmonics describe the portion of
the polarization field which has no handedness associated with it,
while the C harmonics describe the other portion of the field which
does have a handedness (just as with the gradient and curl of
a vector field).

This geometric interpretation leads to an important physical
conclusion. Consider a universe containing only scalar
perturbations, and imagine a single Fourier mode of the
perturbations. The mode has only one direction
associated with it, defined by the Fourier vector $\bf k$; 
since the perturbation is scalar,
it must be rotationally symmetric around this axis.
(If it were not, the gradient of the perturbation 
would define an independent physical direction, which 
would violate the assumption of a scalar perturbation.)
Such a mode can have no physical handedness
associated with it, and as a result, the polarization 
pattern it induces in the microwave background couples
only to the G harmonics. Another way of stating this
conclusion is that primordial density perturbations
produce {\it no} C-type polarization as long as
the perturbations evolve linearly.
This property is very useful for constraining 
or measuring other physical effects, several of
which are considered below.

Finally, just as temperature fluctuations are commonly
characterized by their power spectrum $C_l$, polarization
fluctuations possess analogous power spectra. 
We now have three sets of multipole moments, $a_{(lm)}^{\rm T}$,
$a_{(lm)}^{\rm G}$, and $a_{(lm)}^{\rm C}$,
which fully describe the
temperature/polarization map of the sky. 
Statistical isotropy implies that
\begin{eqnarray}
\left\langle a_{(lm)}^{{\rm T}\,*}a_{(l'm')}^{\rm T}\right\rangle =
                   C^{\rm T}_l \delta_{ll'}\delta_{mm'},&\qquad\qquad&
\left\langle a_{(lm)}^{{\rm G}\,*}a_{(l'm')}^{\rm G}\right\rangle = 
                                   C_l^{\rm G} \delta_{ll'}\delta_{mm'},
                                    \cr
\left\langle a_{(lm)}^{{\rm C}\,*}a_{(l'm')}^{\rm C}\right\rangle =
                   C_l^{\rm C} \delta_{ll'}\delta_{mm'},&\qquad\qquad&
\left\langle a_{(lm)}^{{\rm T}\,*}a_{(l'm')}^{\rm G}\right\rangle =
                                   C_l^{\rm TG}\delta_{ll'}\delta_{mm'},
                                      \cr
\left\langle a_{(lm)}^{{\rm T}\,*}a_{(l'm')}^{\rm C}\right\rangle =
                   C_l^{\rm TC}\delta_{ll'}\delta_{mm'},&\qquad\qquad&
\left\langle a_{(lm)}^{{\rm G}\,*}a_{(l'm')}^{\rm C}\right\rangle =
                                   C_l^{\rm GC}\delta_{ll'}\delta_{mm'},
\label{cldefs}
\end{eqnarray}
where the angle brackets are an average over all realizations of the
probability distribution for the cosmological initial conditions.
Simple statistical estimators of the various $C_l$'s can be
constructed from maps of the microwave background temperature
and polarization.

For Gaussian theories, the statistical properties of a
temperature/polarization map are specified fully by these six
sets of multipole moments.
In addition, the scalar spherical harmonics $Y_{(lm)}$ and the G
tensor harmonics $Y_{(lm)ab}^{\rm G}$ have parity $(-1)^l$, but the C
harmonics $Y_{(lm)ab}^{\rm C}$ have parity $(-1)^{l+1}$.
If the large-scale perturbations in the early universe were invariant
under parity inversion, then
$C_l^{\rm TC}=C_l^{\rm GC}=0$.  
The arguments in the previous paragraph about handedness
further imply that for {\it scalar} perturbations,
$C_l^{\rm C}=0$. A question of substantial theoretical and
experimental interest is what kinds of physics produce
measurable nonzero $C_l^{\rm C}$, $C_l^{\rm TC}$, and
$C_l^{\rm GC}$. This question is addressed in the following section.

\begin{figure}[htbp]
\centerline{\psfig{file=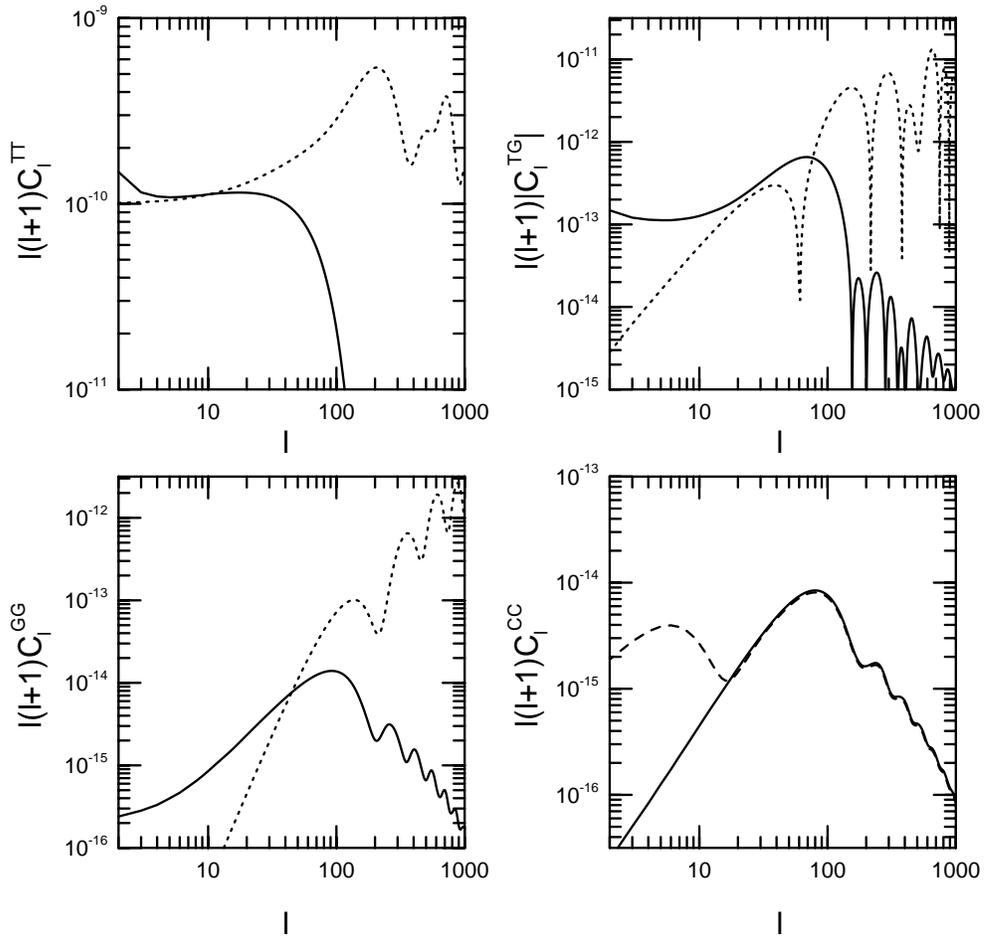,width=7in}}
\bigskip
\caption{
          Theoretical predictions for the four nonzero CMB
	  temperature-polarization spectra as a function
	  of multipole moment $l$.  The solid curves are the
	  predictions for a COBE-normalized scalar perturbations,
          while the dotted curves are 
	  COBE-normalized tensor perturbations. 
	  Note that the panel for $C_l^{\rm C}$ 
          contains no dotted curve since scalar perturbations
	  produce no ``C'' polarization component; instead,
          the dashed line in the lower right panel shows a
	  reionized model with optical depth $\tau=0.1$ to the
	  surface of last scatter. 
       }
\label{fig:clsplot}
\end{figure}

The power spectra can be computed for a given cosmological
model through well-known numerical techniques.
A set of power spectra for scalar and
tensor perturbations in a typical inflation-like cosmological
model, generated with the
CMBFAST code \cite{SelZal96} are displayed in Fig.~\ref{fig:clsplot}.

\section{Polarization and Physical Effects}
\label{physics}

What is microwave background polarization good for? One basic
and model-independent answer to this question was outlined
above: polarization can provide a clean demonstration of the
existence of acoustic oscillations in the early universe. The
fact that three of the six polarization-temperature power spectra
are zero for linear scalar perturbations gives several other
interesting and model-independent probes of physics.

The most important is that the ``curl'' polarization power
spectrum directly reflects the existence of any vector
(vorticity) or tensor (gravitational wave) metric perturbations.
Inflation models generically predict a nearly-scale invariant
spectrum of tensor perturbations, while defects or other
active sources produce significant amounts of both vector
and tensor perturbations. If the measured temperature power
spectrum of the microwave background turns out to look
different than what is expected in the broad class of inflation-like
cosmological models, polarization will tell what part 
of the temperature anisotropies arise from vector and
tensor perturbations. More intriguingly, in inflation models,
the amplitude of the tensor perturbations is directly
proportional to the energy scale at which inflation occurred, so
characterizing the gravitational wave background becomes a
probe of GUT-scale physics at $10^{16}$ GeV! Inflation also
predicts potentially measurable relationships between the
amplitudes and power law indices of the primordial density
and gravitational wave perturbations (see \cite{Lidetal97}
a comprehensive overview), and measuring a
$C_l^{\rm C}$ power spectrum appears to be the only way to
obtain precise enough measurements of the tensor perturbations
to test these predictions. A microwave background map with
forseeable sensitivity could measure gravitational wave
perturbations with amplitudes smaller than $10^{-3}$ times
the amplitude of density perturbations \cite{KamKos98}, 
thanks to the fact
that the density perturbations don't contribute to $C_l^{\rm C}$.
The tensor perturbations generally contribute significantly
to the temperature perturbations at angular scales larger
than two degrees ($l\lesssim 100$) in a flat universe
but have a much broader range of scales in polarization
($50 \lesssim l \lesssim 500$).
For tensor and vector
perturbations, the amplitude of the C-polarization
is generally about the same as that of the G-polarization;
if the perturbations inducing the COBE temperature anisotropies are
10\% tensors, then we expect the peak of
$l^2C_l^{\rm C}\simeq 10^{-15}$ at
angular scales around $l=80$.
An experimental challenge not for the faint of heart! 

A second source of C-type polarization is gravitational
lensing. The mass distribution in the universe between us and
the surface of last scatter will bend the geodesics of the
microwave background photons. This lensing can be described by
an effective displacement field, in which the temperature and
polarization at each point of the sky in an unlensed universe
is mapped to a nearby but different point on the sky when lensing
is accounted for. The displacement alters the shape of temperature
contours in the microwave background, and likewise distorts
the polarization pattern, inducing some curl component to the
polarization field. Detailed calculations of this
effect and the induced $C_l^{\rm C}$ have been made by
\citeasnoun{ZalSel98}. The amplitude of this effect
is expected to be around $l^2C_l^{\rm C}\simeq 10^{-14}$ on a broad
range of subdegree
angular scales ($200\lesssim l \lesssim 3000$) with the power spectrum
peaking around $l=1000$ in a flat universe. This lensing polarization 
signal is just at the limit of detectability for the upcoming
Planck satellite; future polarization satellites with better
sensitivity could make detailed lensing maps based on the curl
component of microwave background polarization.
It is interesting to note that tensor perturbations and gravitational
lensing are substantially distinguishable by their different
angular scales. Note that the most recent version of the
publicly available CMBFAST code by
Seljak and Zaldarriaga \cite{SelZal96} computes polarization from both
tensor modes and from gravitational lensing.

A third source of C-type polarization is a primordial magnetic
field. If a magnetic field was present at recombination,
the linear polarization of electromagnetic radiation would
undergo a Faraday rotation as it propagated through the
surface of last scatter while significant numbers of free
electrons were still present. (Such rotation could also
occur after reionization, but both the electron density
and the field strength would be much smaller and the
resulting rotation is small compared to the primordial signal).
This effect rotates an initial G-type polarization field
into a C-type polarization field. A detailed estimate of
the magnitude of this effect \cite{KosLoe97} shows
that a primordial field with present strength $10^{-9}$ gauss
induces a measurable one-degree rotation in the polarization
at a frequency of 30 GHz. Faraday rotation depends quadratically
on wavelength of the radiation, so down at 3 GHz, the rotation would
be a huge 100 degrees (although the polarized emission from
synchrotron radiation would also be correspondingly larger).
Such a rotation will induce $l^2 C^{\rm C}$ at a level of
between $10^{-15}$ for one degree of rotation and $10^{-11}$
for large rotations. Additionally, it has been pointed out that
Faraday rotation will contribute also to the $C_l^{\rm TC}$
cross-correlation at corresponding levels \cite{ScaFer97}
as well as to $C_l^{\rm GC}$.
Investigation of the angular dependence
and detectability of such a signal is ongoing \cite{MacKos99}.
The best current constraints on a homogeneous component of a primordial
magnetic field come from COBE constraints on anisotropic Bianchi
spacetimes \cite{BarFerSil97}, because a universe which contains
a homogeneous magnetic field cannot be statistically isotropic.
Detection of a significant
primordial magnetic field would both provide the
seed field needed to generate current galactic and subgalactic-scale
magnetic fields via the dynamo mechanism, and also provide a very
interesting constraint on fundamental particle physics, particularly
if a field on large scales is detected (see, e.g., 
\citeasnoun{TurWid88} or \citeasnoun{GasGioVen95}).

Faraday rotation from magnetic fields is a special case
of cosmological birefringence: rotation of
polarization by differing amounts depending on direction
of observation. Such rotation could arise from 
interactions between photons and other unknown fields.
Constraints on the C-polarization of the microwave background
could strongly constrain new pseudoscalar particles
(see, e.g., \citeasnoun{CarFie97}).
More generally, non-zero cosmological contributions to
the $C_l^{\rm TC}$ and $C_l^{\rm GC}$ cross correlations,
which must be zero if parity is a valid symmetry of
the cosmological perturbations, would indicate some
intrinsic parity to either the primordial perturbations
\cite{LueWanKam98}
or to some interaction of the microwave background photons
\cite{Car98}. These types of effects are generally
independent of photon frequency, so they can be distinguished
from Faraday rotation through microwave background frequency
dependence.

The above signals are all model-independent probes of new physics
using microwave background polarization. An additional less
daring but initially more useful and important use of polarization
is in determining and constraining the basic background
cosmology of the universe. It has been appreciated for
several years now that the microwave background offers the
cleanest and most powerful constraint on the gross features
of the universe \cite{Junetal96}. If the universe is
described by an inflation-type model, with nearly scale-invariant
initial adiabatic perturbations which evolve via gravitational
instability, then the power spectrum of microwave background
temperature fluctuations can strongly constrain nearly all
cosmological parameters describing the universe: densities
of various matter and energy components, amplitudes and
power laws of initial density and gravitational wave perturbations,
the Hubble parameter, and the redshift of reionization. More
recent work \cite{ZalSelSpe97,EisHuTeg99} has shown that
the addition of polarization information can help tighten
these constraints considerably, mainly because the new information
now gives four theoretical power spectra to match
instead of just one.
Polarization particularly helps constrain the reionization
redshift and the baryon density \cite{ZalHar95}.
Polarization will also be important for deciphering the
universe if measurements of the temperature anisotropies reveal
that the universe is {\it not} described by the simple
class of inflation-like cosmological models: it is a strong
discriminator between vector and tensor perturbations and
scalar perturbations \cite{KamKosSte97}.

Finally, no discussion of this sort would be completely
honest without mentioning the thorny issue of foreground emission.
We are gradually concluding that foregrounds have some
non-negligible effect on temperature anisotropies, but that
the amplitudes of various foregrounds are small enough that
they will not substantially hinder our ability to draw
cosmological conclusions from microwave background temperature
maps (see, e.g., \citeasnoun{Teg98} for a recent estimate). 
Whether the same will prove true for polarization is
unknown at present. Free-free emission is likely to have only
negligible polarization, but synchrotron emission will be
strongly polarized, and the polarization of dust emission is
difficult to estimate reliably \cite{DraLaz98}. 
Polarized emission from
radio point sources is another potential problem.
No present measurements have
had sufficient sensitivity to detect polarized emission from
any of these foreground sources, so it is difficult to predict
the foreground impact. My own guess is that the G-polarization
component, from which acoustic oscillations can be confirmed and
from which parameter estimation can be significantly improved,
will face foreground contamination comparable to 
the temperature anisotropies. If so, and if the polarization
foregrounds are divided evenly between C and G polarization
components, then control of foregrounds will become crucial
for the very interesting physics probed by the cosmological
C polarization.  But I fully
expect that through a combination of techniques, including
carefully tailored sky cuts, measurements at many frequencies,
improved theoretical understanding,
foreground nongaussianity, and
foreground template matching,
we will separate out the small cosmological polarization
signals from whatever polarized foregrounds are out there.

The next five years will bring us microwave background temperature
maps of vastly improved sensitivity and resolution, and almost
certainly the first detection of microwave background polarization.
These observations will provide us with very tight constraints
on our cosmological model, or else will reveal some new and
unexpected aspect of our universe. Either way, the microwave
background will be the cornerstone of a mature cosmology. What is
left to do after Planck? One good answer to this question, I believe, 
is very high
sensitivity measurements of microwave background polarization.
Such observations hold the promise of probing the potential
driving inflation, detecting primordial magnetic fields,
mapping the matter distribution in the universe, and likely
a variety of other interesting physics yet to be explored.

This work has been supported by the NASA Theory Program.
Portions of this work were
done at the Institute for Advanced Study.

\end{document}